\documentclass[draftclsnofoot,12pt, onecolumn]{IEEEtran}
\usepackage{amsmath}
\usepackage{epsfig}
\usepackage{amssymb}
\usepackage{latexsym}
\usepackage{graphicx}
\usepackage{cite}

\begin{document}
\title{Solutions to the Incomplete Toronto Function and Incomplete Lipschitz-Hankel Integrals}
\author{\IEEEauthorblockN{Paschalis C. Sofotasios\\}
\IEEEauthorblockA{School of Electronic and Electrical Engineering\\
University of Leeds, UK\\
e-mail: p.sofotasios@leeds.ac.uk\\}
\and
\IEEEauthorblockN{Steven Freear\\}
\IEEEauthorblockA{School of Electronic and Electrical Engineering\\
University of Leeds, UK\\
e-mail: s.freear@leeds.ac.uk}}
\maketitle
\begin{abstract} 
This paper provides novel analytic expressions for the incomplete Toronto function, $T_{B}(m,n,r)$, and the incomplete Lipschitz-Hankel Integrals of the modified Bessel function of the first kind, $Ie_{m,n}(a,z)$. These expressions are expressed in closed-form and are valid for the case that $n$ is an odd multiple of $1/2$, i.e. $n \pm 0.5\in\mathbb{N}$. Capitalizing on these, tight upper and lower bounds are subsequently proposed for both $T_{B}(m,n,r)$ function and $Ie_{m,n}(a,z)$ integrals. Importantly, all new representations are expressed in closed-form whilst the proposed bounds are shown to be rather tight. To this effect, they can be effectively exploited in various analytical studies related to wireless communication theory. Indicative applications include, among others, the performance evaluation of digital communications over fading channels and the information-theoretic analysis of multiple-input multiple-output systems.
\end{abstract}


\section{Introduction}
\indent
Special functions are undoubtedly an inevitable mathematical tool in almost all areas of natural sciences and engineering. In the wide field of telecommunications, their use in numerous analytical studies often renders possible the derivation of explicit expressions for important performance metrics such as channel capacity and error probability. Furthermore, their corresponding computation is typically not considered laborious since the majority of them are included as built-in functions in widely known software packages such as $Maple$, $Mathematica$ and $Matlab$. Based on this, both the algebraic representation and computational realization of any associated analytical expressions have been undoubtedly simplified. \\
\indent
The incomplete Toronto function and the incomplete Lipschitz-Hankel integrals (ILHIs) appear, among others, in various analytic solutions of problems related to wireless communications. They were both proposed a few decades ago and they are denoted as $T_{B}(m,n,r)$ and $Ze_{m,n}(a,z)$, respectively \cite{B:Abramowitz}. The incomplete Toronto function constitutes a special case of the complete Toronto function, which was initially proposed by Hatley in \cite{J:Hatley}. It also includes as a special case the Marcum Q-function and has been used in studies related to statistics, signal detection and estimation, radar systems and error probability analysis, \cite{J:Fisher, J:Marcum, J:Swerling}. Its definition is typically given in integral form  which involves an arbitrary power term, an exponential term and a modified Bessel function of the first kind while alternative representations include two infinite series, \cite{J:Sagon}. In the same context, the ILHIs belong to a class of incomplete cylindrical functions that have been largely encountered in analytical solutions of numerous problems in electromagnetics, \cite{B:Maksimov, J:Dvorak} and the references therein. In communication theory, they have been used in recent investigations related with the error rate analysis of MIMO systems under imperfect channel state information (CSI) employing adaptive modulation, transmit beamforming and maximal ratio combining (MRC), \cite{J:Paris}. \\ 
\indent
However, in spite of the evident importance of the $T_{B}(m,n,r)$ functions and the $Ze_{m,n}(a,z)$ integrals, they are both neither tabulated, nor included as built-in functions in the aforementioned popular software packages. As a consequence, they appear inconvenient to handle both analytically and computationally. Motivated by this, the aim of this work is the derivation of novel analytic results for $T_{B}(m, n, r)$ and $Ze_{m, n}(a, z)$. In more details, explicit expressions and upper and lower bounds to the $T_{B}(m,n,r)$ function and $Ie_{m,n}(a,z)$ integrals\footnote{Only the $I_{n}(x)$ function is considered in the present work.}, are derived for the case of $n + 0.5 \in \mathbb{N}$. The offered results are expressed in closed-form and have a tractable algebraic representation which ultimately renders them useful for utilization in various analytical studies associated to wireless communications. Indicatively, such studies include, among others, the derivation of explicit expressions for important performance measures, such as channel capacity and probability of error, in the wide field of digital communications over fading channels and the information-theoretic analysis of MIMO systems, among others \cite{Add_4, Add_5, Add_6, Add_7, Add_8}. \\
\indent
The remainder of this paper is structured as follows: Section II revisits the definition and basic principles of the $T_{B}(m, n, r)$ function and the $Ze_{m, n}(a, z)$ integrals. Subsequently, Sections III and IV are devoted to the derivation of novel expressions and upper and lower bounds, respectively. Finally, discussion on the potential applicability of the offered relationships in wireless communications along with closing remarks, are provided in Section V. 
%
%
\section{Definitions and existing representations}
\subsection{The Incomplete Toronto Function}
\indent
The incomplete Toronto function is defined as,

\begin{equation} \label{1} 
T_{B}(m,n,r) \triangleq 2r^{n-m+1} e^{-r^{2}} \int_{0}^{B} t^{m-n} e^{-t^{2}}I_{n}(2rt) dt
\end{equation}
where $I_{n}(.)$ denotes the modifies Bessel function of the first kind and order $n$. For the special case that $n = (m-1)/2$, it is equivalently expressed in terms of the Marcum Q-function, $Q_{m}(a,b)$, as follows

\begin{equation} \label{2} 
T_{B}\left(m,\frac{m-1}{2},r\right) = 1 - Q_{\frac{m+1}{2}}\left(r\sqrt{2},B\sqrt{2}\right)
\end{equation}
Two alternative representations for the $T_{B}(a,b,r)$ function, in the form of infinite series, were reported in \cite{J:Sagon}, namely, 

\begin{equation} \label{3} 
T_{B}(m,n,r) = \frac{B^{2a}r^{2(n-a+1)}}{n!}e^{-B^{2}-r^{2}}\sum_{k=0}^{\infty}\frac{r^{2k}Y_{k}}{(a)_{k+1}}
\end{equation}
and 

\begin{equation} \label{4} 
T_{B}(m,n,r) = r^{2(n-a+1)}e^{-r^{2}}\sum_{k=0}^{\infty} \frac{r^{2k}\gamma(a+k, B)}{k!(n+k)!}
\end{equation}
where, 

$$
Y_{k} = \sum_{i=0}^{k}\frac{(a)_{i}r^{2i}}{(n+1)_{i} i!} 
$$
and $ a = (m+1)/ 2$. \\
\indent
The notations $(a)_{k}$ and $\gamma(c,x)$ denote the Pochhammer symbol and the lower incomplete gamma function, respectively, \cite{B:Tables, Add_1, Add_2, Add_3}. Notably, equations \eqref{3} and \eqref{4} are exact while, their algebraic representation appears to be relatively tractable. Nevertheless, their infinite form raises convergence and truncation issues.

\subsection{The Incomplete Lipschitz-Hankel Integrals}
 
The general ILHI is defined as, 

\begin{equation}  \label{5} 
Ze_{m,n}(a,z) \triangleq \int_{0}^{z}x^{m}e^{-ax}Z_{n}(x)dx
\end{equation}
where $m$, $n$, $a$, $z$ may be also complex \cite{B:Maksimov, J:Dvorak}. The notation $Z_{n}(x)$ denotes one of the cylindrical functions  $J_{n}(x)$, $I_{n}(x)$, $Y_{n}(x)$, $K_{n}(x)$, $H_{n}^{1}(x)$ or $H_{n}^{2}(x)$, \cite{B:Abramowitz}. An alternative representation for the ILHIs of the first-kind modified Bessel functions, was recently reported in \cite{J:Paris}. This representation is given in terms of the Marcum Q-function and is expressed as

$$
Ie_{m,n}(a,z) = A_{m,n}^{0}(a) + e^{-ax} \sum_{i=0}^{m}\sum_{j=0}^{n+1}B_{m,n}^{i,j}(a)x^{i}I_{j}(x) 
$$
\begin{equation} \label{6} 
+ A_{m,n}^{1}(a)Q_{1}\left(\sqrt{\frac{x}{a+\sqrt{a^{2}-1}}}, \sqrt{x}\sqrt{a+\sqrt{a^{2}-1}} \right)
\end{equation}
where the set of coefficients $A_{m,n}^{l}(a)$ and $B_{m,n}^{i,j}(a)$ can be obtained recursively. As already mentioned, the above relationship has been shown to be useful in the error rate analysis of MIMO systems under imperfect channel state information (CSI). 
%
%
\section{An Exact Representation and Bounds for the Incomplete Toronto Function}
\indent 
Recalling Section I, the $T_{B}(m,n,r)$ function is neither expressed in terms of other special and/or elementary functions, nor is it included as built-in function in popular mathematical software packages. Motivated by this, a novel closed-form expression is derived for the case that $n$ is an odd multiple of $1/2$. Capitalizing on this expression, novel closed-form upper and lower bounds are subsequently deduced. 
\subsection{A Closed-Form Solution for the $T_{B}(m,n,r)$ Function.}
$ $\\
\noindent
\textbf{Theorem 1.} \textit{For $m, r ,B \in \mathbb{R^{+}} $, $n \pm \frac{1}{2} \in \mathbb{N}$, $m \pm \frac{1}{2} \in \mathbb{N}$  and  $m\geq n$,  the following relationship holds,}

$$
T_{B}(m,n,r) = \sum_{k=0}^{n - \frac{1}{2}} \sum_{l=0}^{L-k}\frac{\left(n+k-\frac{1}{2} \right)!\, (L-k)!\,2^{-2k}r^{-2k-l}}{\sqrt{\pi}k! \left(n-k-\frac{1}{2} \right)!l!(L-l-k)!} \times 
$$
\begin{equation} \label{7} 
\left\lbrace (-1)^{m-k-l} \frac{\gamma\left(\frac{l+1}{2}, (B+r)^{2} \right)}{2}     - (-1)^{k} \frac{\gamma\left(\frac{l+1}{2}, (B-r)^{2} \right)}{2}             \right\rbrace
\end{equation}
\textit{where $L = m-n-\frac{1}{2}$ and $\gamma(a,x)$ denotes the lower incomplete gamma function} \cite{B:Abramowitz}. 
\\
$ $ \\
\noindent
\textbf{Proof.} By setting in $x = 2rt$ and assuming $ n + \frac{1}{2} \in \mathbb{N}$, the corresponding $I_{n}(x)$ function can be re-written according to \cite[eq. (8.467)]{B:Tables}, namely,

\begin{equation} \label{8} 
I_{n}(2rt) = \sum_{k=0}^{n - \frac{1}{2}} \frac{\left(n+k-\frac{1}{2} \right)! \left[ (-1)^{k} e^{2rt} + (-1)^{n+\frac{1}{2}}e^{-2rt}\right]}{k!\sqrt{\pi} \left(n - k - \frac{1}{2} \right)! 2^{2k+1}t^{k+\frac{1}{2}}r^{k + \frac{1}{2}}} 
\end{equation}
By substituting in \eqref{1} and making use of the basic identity: $(a \pm b)^{2} = a^{2} \pm 2ab +b^{2}$, one obtains

$$
T_{B}(m,n,r) = \sum_{k=0}^{n - \frac{1}{2}} \frac{(n+k - \frac{1}{2})! r^{n-m-k+\frac{1}{2}}}{k!\sqrt{\pi}(n-k-\frac{1}{2})!2^{2k}} \times
$$
\begin{equation} \label{9} 
\left\lbrace (-1)^{k} \int_{0}^{B} t^{L} e^{-(t-r)^{2}}dt + (-1)^{n+\frac{1}{2}} \int_{0}^{B} t^{L} e^{-(t+r)^{2}} dt \right\rbrace
\end{equation}
where $L=m-n-k-\frac{1}{2}$. Evidently, a closed-form solution to the above expression is subject to evaluation of the two involved integrals. To this end, with the aid of \cite[eq. (1.3.3.18)]{B:Prudnikov}, for the case that $ L=2k$ with $k \in \mathbb{N}$, one obtains

\begin{equation}
\int_{0}^{B}x^{L}e^{-(x+a)^{2}}dx + \int_{0}^{B}x^{L}e^{-(x-a)^{2}}dx= \sum_{l=0}^{L} \frac{L! a^{L-l}}{l!(L-l)!} \left[ (-1)^{L-l}\int_{0}^{B+a}x^{l}e^{-x^{2}}dx + \int_{0}^{B-a}x^{l} e^{-x^{2}} dx \right]
\end{equation}
To this effect, equation (9) can be re-written as follows

$$
T_{B}(m,n,r) = \sum_{k=0}^{n - \frac{1}{2}} \sum_{l=0}^{L}\frac{\left(n+k-\frac{1}{2} \right)!\, L!\,2^{-2k}r^{-2k-l}}{\sqrt{\pi}k! \left(n-k-\frac{1}{2} \right)!l!(L-l)!} \times
$$
\begin{equation} \label{10} 
\left\lbrace (-1)^{m-k-l}\int_{0}^{B+r}t^{l}e^{-t^{2}}dt - (-1)^{k}\int_{0}^{B-r}t^{l}e^{-t^{2}}dt \right\rbrace
\end{equation}

Evidently, the above integrals can be solved in terms of the lower incomplete gamma function according to \cite[eq. (3.381.3)]{B:Tables}. Therefore, equation \eqref{7} is finally deduced and the proof is completed. $\blacksquare$ 
\subsection{Upper and Lower Bounds for the $T_{B}(m,n,r)$ Function}
\indent
Novel bounds to the incomplete Toronto function may be straightforwardly deduced from Theorem $1$. \\
$ $\\
\noindent
\textbf{Corollary 1:} \textit{For $m, r, B, b \in \mathbb{R^{+}} $ and  $m\geq n$, the following inequality holds,}

\begin{equation} \label{11} 
T_{B}(m,n,r) \, > \, T_{B}\left(m,  \lceil n + 0.5 \rceil - 0.5, r \right)
\end{equation}
\textit{where $T_{B}\left(m,  \lceil n + 0.5 \rceil - 0.5, r \right)$ is given in closed-form in \eqref{7} since it meets the condition $n + \frac{1}{2} \in \mathbb{N}$.}
\\
$ $ \\
\noindent
\textbf{Proof:} The incomplete Toronto function is strictly decreasing with respect to $n$. To this effect, for an arbitrary real positive value $a$, it follows that $T_{B}(m,n+a,r)<T_{B}(m,n,r)$. As a result, by recalling that the incomplete Toronto function can be expressed in closed-form for $n + \frac{1}{2} \in \mathbb{N}$, by upper ceiling the $T_{B}(m,n,r)$ according to the identity $ \lceil a + 0.5 \rceil - 0.5 > a$, lower bounds the function as in (11) and thus, the proof is completed. $\blacksquare$\\
$ $\\
\noindent
\textbf{Corollary 2:} \textit{For $m, r, B, b \in \mathbb{R^{+}} $ and  $m\geq n$, the following inequality holds,}

\begin{equation} \label{12} 
T_{B}(m,n,r) \, < \, T_{B}\left(m,  \lfloor n - 0.5 \rfloor + 0.5, r \right)
\end{equation}
\textit{where $T_{B}\left(m,  \lfloor n - 0.5 \rfloor + 0.5, r \right)$ is given in closed-form in \eqref{7}  is given in closed-form in \eqref{7} since it meets the condition $n + \frac{1}{2} \in \mathbb{N}$.}
\\
$ $ \\
\noindent
\textbf{Proof:} The proof follows immediately from Theorem $1$ and Corollary $1$. $\blacksquare$ 
\begin{figure}[h]
\centerline{\psfig{figure=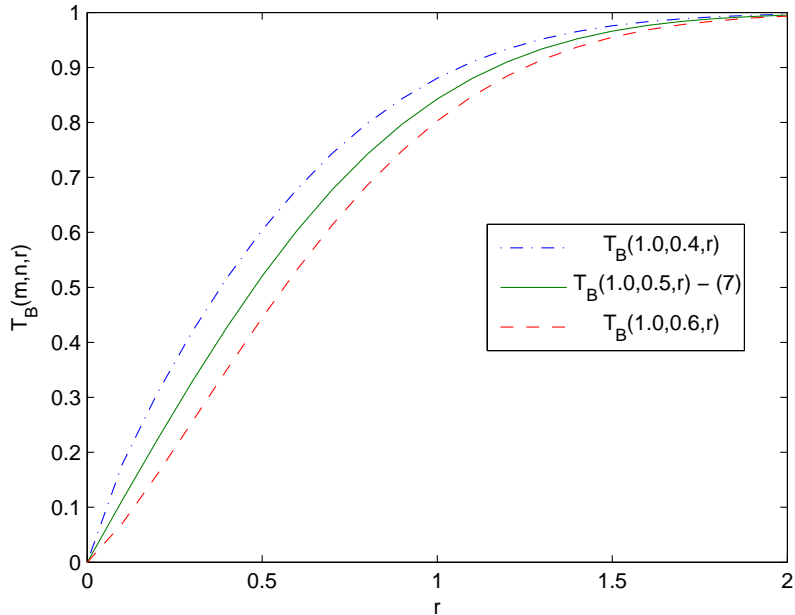, height=9cm, width=12cm}}
\caption{Behaviour of the exact solution and the performance bounds to $T_B(m,n,r)$ for $m=1.0$ and different $n$}
\end{figure}
\subsection{Numerical Results} 
\indent
The validity of the derived closed-form expression and the general behaviour of the offered bounds are illustrated in Figures $1$ and $2$. Specifically, the behaviour of \eqref{7} with respect to $r$, is depicted for Figure $1$ for $n=0.5$ and $m=1.0$.  Results obtained from numerical integrations for $m=1.0$ and $n=0.4$, $n=0.5$ and $n=0.6$ are also demonstrated for comparative purposes. In the same context, equation \eqref{7} is depicted in Figure $2$ for $m=3.0$ and $n=2.5$ along with numerical results for the cases that $n=2.4$, $n=2.5$ and $2.6$. Evidently, one can observe that \eqref{7} is in exact agreement with the corresponding numerical results while the overall tightness of the derived bounds is shown to be quite adequate over the whole range of values of $r$. 
%
%
\section{An Exact Representation and Bounds for the the Incomplete Lipschitz-Hankel Integrals}
\indent 
Likewise the $T_{B}(m,n,r)$ function, the ILHIs are neither tabulated, nor are they built-in in widely known mathematical software packages. However, their algebraic form constitutes possible the derivation of a closed-form expression for the case that $n$ is an odd multiple of $1/2$. \\
\subsection{A Closed-Form Solution for the $Ie_{m,n}(z;a)$ Integrals}
$ $ \\
\noindent
\textbf{Theorem 2:} \textit{For $m, r, B \in \mathbb{R^{+}} $, $n + 0.5 \in \mathbb{N}$ and $m\geq n$, the following closed-form relationship holds,}

\begin{equation} \label{13} 
Ie_{m,n}(a, z) = \sum_{k=0}^{n - \frac{1}{2}} \frac{\left(n+k-\frac{1}{2} \right)!}{\sqrt{\pi}k!\left(n-k-\frac{1}{2} \right)!2^{k+\frac{1}{2}}}  \left\lbrace (-1)^{k} \frac{\gamma \left(P, (a-1)z \right) }{(a-1)^{P}} + (-1)^{n + \frac{1}{2}} \frac{\gamma \left(P, (a+1)z \right) }{(a+1)^{P}} \right\rbrace
\end{equation}
where 

 \begin{equation}
 P=m-k+\frac{1}{2}. 
  \end{equation} 
\\
$ $ \\
\noindent
\textbf{Proof:} By expressing the $I_{n}(x)$ function in with its closed-form representation according to \cite[eq. (8.467)]{B:Tables} and substituting in  \eqref{5}, it immediately follows that

\begin{figure}[h]
\centerline{\psfig{figure=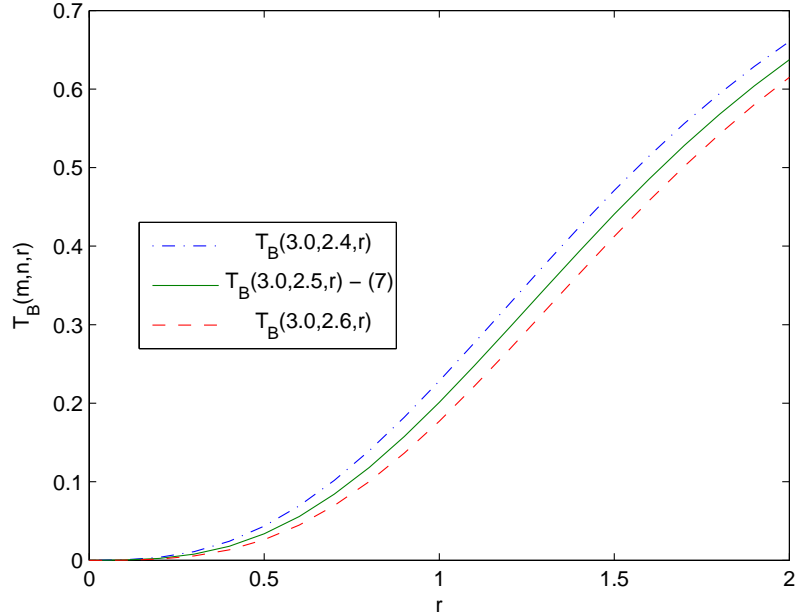, height=9cm, width=12cm}}
\caption{Behaviour of the exact solution and the performance bounds to $T_B(m,n,r)$ for $m=3.0$ and different $n$}
\end{figure}
 
\begin{equation} \label{14} 
Ie_{m,n}(a, z) = \sum_{k=0}^{n-\frac{1}{2}} \frac{\left(n+k- \frac{1}{2}\right)! 2^{-k - \frac{1}{2}}}{\sqrt{\pi} k!\left(n-k-\frac{1}{2} \right)!}   \left\lbrace (-1)^{k}\int_{0}^{z}x^{P}e^{-ax}e^{x}dx + (-1)^{n + \frac{1}{2}} \int_{0}^{z} x^{P}e^{-ax}e^{-x}dx \right\rbrace
\end{equation}
The involved integrals in \eqref{14} have the form of the lower incomplete gamma function. Hence, by carrying out some necessary algebraic manipulations and with the aid of \cite[eq. (3.381.3)]{B:Tables}, one obtains \eqref{13}, which completes the proof. $\blacksquare$ \\
$ $\\
\noindent
\textbf{Remark:} The present analysis was limited in the consideration of the $I_{n}(x)$ function. Nevertheless, similar expressions can be analogously derived for the case of the Bessel functions $J_{n}(x)$, $Y_{n}(x)$, $K_{n}(x)$ as well as the Hankel functions, $H_{n}^{(1)}(x)$ and $H_{n}^{(2)}(x)$, \cite{B:Abramowitz}. 
\subsection{Upper and Lower bounds for the $Ie_{m,n}(a, z)$ Integrals}
$ $ \\
\noindent
\textbf{Corollary 3:} \textit{For $m, r, B, n \in \mathbb{R^{+}}$ and $m\geq n$, the following inequality holds}

\begin{equation} \label{15} 
I_{m,n}(a, z) > I_{m,\lceil n + \frac{1}{2}\rceil - \frac{1}{2} }(a, z)
\end{equation}
\textit{where $I_{m,n - \frac{1}{2}}(a, z)$ can be expressed in closed-form according to (13) since it always meet the condition $n + \frac{1}{2} \in \mathbb{N}$.  }
\\
$ $ \\
\noindent
\textbf{Proof:} it is noted that the $I_{m,n}(a, z)$ integrals are monotonically decreasing with respect to $n$. Thus, for an arbitrary real positive value $a$, $a \in \mathbb{R^{+}}$, it follows that $I_{m+a,n}(a, z) < I_{m,n}(a, z)$. Thus, by recalling the property $\lceil n + 0.5\rceil - 0.5>n$ the closed-form lower bound in \eqref{15} is deduced. $\blacksquare$\\

$ $\\
\noindent
\textbf{Corollary 4:}  \textit{For $m, r, B, n \in \mathbb{R^{+}}$ and $m\geq n$, the following inequality holds}
\begin{equation} \label{16} 
I_{m,n}(a, z) < I_{m,\lfloor n - \frac{1}{2}\rfloor + \frac{1}{2} }(a, z)
\end{equation}
\textit{where $I_{m,\lfloor n - \frac{1}{2}\rfloor + \frac{1}{2} }(a, z)$ can be expressed in closed-form according to (13) since it always meet the condition $n + \frac{1}{2} \in \mathbb{N}$.  }
\\
$ $ \\
\noindent
\textbf{Proof:} The proof follows immediately from Theorem $2$ and Corollary $3$. $\blacksquare$ \\
\begin{figure}[h]
\centerline{\psfig{figure=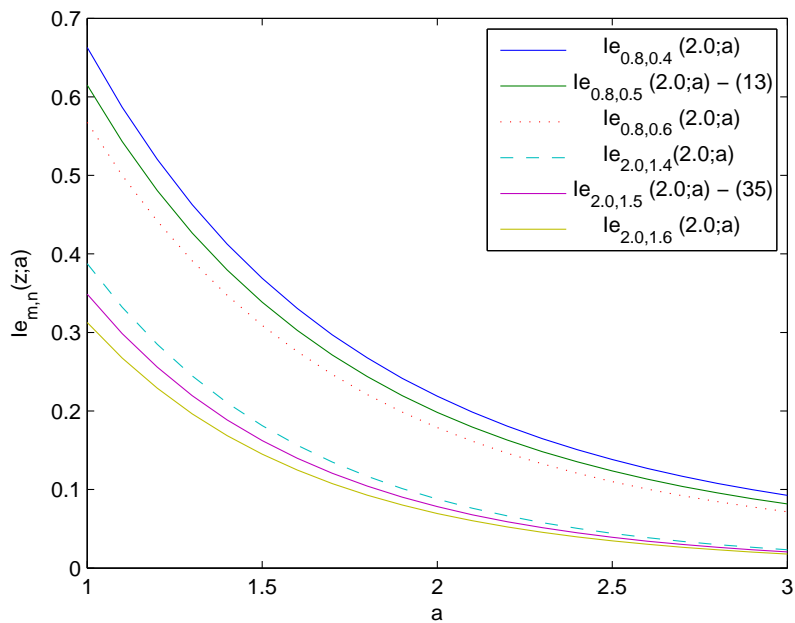, height=9cm, width=12cm}}
\caption{Behaviour of the exact solution and the performance bounds to $Ie_{m,n}(a, z)$ for different values of $n$ and $m$}
\end{figure}
\subsection{Numerical Results}
\indent
The validity and behaviour of the offered results are demonstrated in Figure $3$. One can observe the exactness of \eqref{13} along with the evident tightness of the proposed bounds. Importantly, the achieved tightness holds over the whole range of parametric values. 
\section{Conclusion}
\indent 
In this work, explicit representations and upper and lower bounds were derived for the incomplete Toronto function and the incomplete Lipschitz-Hankel integrals of the modified Bessel function of the first kind. The offered results are novel and are expressed in closed-form. This is sufficiently advantageous since it renders them suitable for application in various studies relating to the performance analysis of digital communications over fading channels, among others. 
$ $\\
\bibliographystyle{IEEEtran}

\end{document}